\documentclass[letter]{aa} % for the letters

\usepackage{natbib}
\usepackage{color}
\usepackage{hyperref}
\hypersetup{colorlinks=true,allcolors=[rgb]{0,0,0.8}}

\newcommand{\kms}{km~s$^{-1}$}
\newcommand{\mum}{$\mu$m}

\usepackage{graphicx}
\usepackage{placeins}
\usepackage{float}

\makeatletter
\renewcommand*\aa@pageof{, page \thepage{} of \pageref*{LastPage}}
\makeatother

\usepackage{threeparttable}
\usepackage{txfonts}

\begin{document}

   \title{ASASSN-21js: A multi-year transit of a ringed disc}

   \author{T. H. Pramono
          \inst{1}
          \and
          M. A. Kenworthy\inst{1}
          \and
          R. van Boekel\inst{2}
          }
   \institute{Leiden Observatory, Leiden University, PO Box 9513, 2300 RA Leiden, The Netherlands
              \and
    Max-Planck Institut f\"{u}r Astronomie, K\"{o}nigstuhl 17, D-69117 Heidelberg, Germany\\
              \email{kenworthy@strw.leidenuniv.nl}
             }
   \date{Received 08 April 2024; accepted 04 July 2024}

% \abstract{}{}{}{}{}
% 5 {} token are mandatory
 
  \abstract
  {}  
   {The early-type star ASASSN-21js started to fade in 2021, as was detected by the All Sky Automated Survey for Supernovae, undergoing a multi-year eclipse that is still underway.
   We interpret this event as being due to a structured disc of material transiting in front of the star.
   The disc is in orbit around a substellar object with the mass and luminosity of a brown dwarf or smaller.
   We want to determine the expected duration and ending date of the eclipse.}
   {We modelled a tilted and inclined azimuthally symmetric ring system around an unseen companion and calculated the resulting time-varying light curve as the object transited in front of the star.
   We made an initial estimate of the ring parameters and used these as inputs to an MCMC algorithm to determine the geometric properties of the rings with associated uncertainties.}
   {The model most consistent with the light curve to date is a two-ring system at high inclination with respect to the line of sight that has a semi-major axis of 71.6 stellar radii.
   With an estimate of the stellar radius, the transverse velocity is around 0.7 \kms{}, which if bound to the star is an orbit with a semi-major axis of around 13000 au, placing it in the Oort cloud of the parent star.
   The transit is ongoing and will finish around MJD 61526 (May 1 2027).
   We encourage the community to continue observing this object in order to understand its properties.}
   {}

   \keywords{giant planet formation --
               moon formation}

   \maketitle
%
%________________________________________________________________

\section{Introduction}\label{sec:intro}

Discs and rings of material have been detected orbiting astrophysical objects, from supermassive black holes down the mass spectrum to minor planets.
The size and substructures contain information about both the dynamics within the disc (indicating the presence of satellites) and external factors that have sculpted it (formation history and local environment).

The advent of multi-year wide-field surveys for the detection of planets that transit their parent star, first from the ground (e.g. HAT, KELT, SuperWASP, ASASSN, and OGLE) and later in space (CoRoT, \textit{Kepler}, TESS, and PLATO) has paved the way for the discovery of extended disc eclipses \citep[e.g. DESK; ]{Rodriguez15} in addition to their primary science cases.
Ever since the detection of an $\sim$ 2 year eclipse of the star $\epsilon$ Auriage \citep{Carroll91}, long-duration ($>$ 1 yr) eclipses by companions with discs have been detected at an ever-increasing rate: TYC 2505-672-1 \citep{Lipunov16,Rodriguez16} has a 3.5 year long and deep eclipse, along with VVV-WIT-08 \citep{Smith21} and several others from the VVV identified in \citet{Lucas24}.
More recently, Gaia triggers have produced two extended eclipsers: Gaia17bpp \citep{Tzanidakis23} and Gaia21bcv \citep{Hodapp24}. The latter exhibits substructure similar to that seen in J1407b \citep{Mamajek12} and PDS 110 \citep{Osborn17} that hints at the presence of exomoons \citep{Kenworthy_2015}.

The All Sky Automated Survey for SuperNovae \citep[ASAS-SN; ][]{Shappee14,Kochanek17} reported that a main-sequence star started to undergo a deep eclipse with a maximum depth of 20 percent. Subsequently assigned the name ASASSN-21js, this eclipse shows changes in its light curve indicative of substructure within the main eclipse event.

In Section~\ref{sec:lightcurve}, we describe the initial report of the stellar dimming and we derive the star's properties from archival photometry. In Section~\ref{sec:model_building}, we rule out intrinsic causes for the variability and perform an initial estimate of the velocity and size of the occulting structure, then argue that a model with two concentric rings of material presents a good fit to the photometry obtained so far. Based on these initial estimates, we refine our initial fit and use MCMC to optimise the parameters of our fit and provide related estimates on the parameters of the two rings and the expected photometry to the end of the eclipse. After a discussion about the derived eclipse properties, we conclude with the predicted light curve and place ASASSN-21js in a broader context.

\section{Properties of the star}\label{sec:star}

ASASSN first detected a dip in the observed flux of a star on June 3 2021, which was then reported through its transient detection page\footnote{\url{https://www.astronomy.ohio-state.edu/asassn/transients.html}} \citep{shappee_man_2014}.
The automatic cross-referencing of this star, now labelled as ASASSN-21js, is identified in Gaia as Gaia DR3 5334823481651325440 (2MASS 11471176-6210367) with a mean magnitude of G = 12.8 mag.

\begin{table}
\caption{Properties of ASASSN-21js}              % title of Table
\label{table:asassn-21js-param}      % is used to refer this table in the text
\centering                                      % used for centering table
\begin{tabular}{l c c}          % centered columns (4 columns)
\hline\hline                        % inserts double horizontal lines
\textbf{Property} & \textbf{Value} & \textbf{Source} \\    % table heading
\hline                                   % inserts single horizontal line
 $\alpha_{ICRS}$(J2015) [$^\circ$] & 176.79897331752 & 1 \\
 $\delta_{ICRS}$(J2015) [$^\circ$] & -62.17689025155 & 1 \\
 Parallax [mas] & $0.3565 \pm 0.0130$ & 1 \\
 Distance [pc] & $2805 \pm 102$ & 1 \\
    $T_{\mathrm{eff}}$ [K] & $14800^{+4200}_{-2800}$ & this work \\
    $\log g$ [cm/s$^2$] & $4.10^{+0.03}_{-0.05}$ & this work \\
    $A_V$ [mag] & $1.38^{+0.10}_{-0.17}$ & this work \\
    $R_*$ [R$_\sun$] & $3.16^{+0.31}_{-0.26}$ & this work \\
    $M_*$ [M$_\sun$]& $4.58^{+1.38}_{-1.16}$ & this work \\
    $L_*$ [L$_\sun$]& $430^{+974}_{-272}$ & this work \\
    \hline     
\end{tabular}
    \tablebib{
    (1)~\citet{gaia_collab, gaia_dr3}. 

    }
\end{table}

\label{sec:SED_fit}

\begin{figure}
    \hspace{-2mm}
    \includegraphics[width=1.0\columnwidth,trim=15 10 24 0,clip]{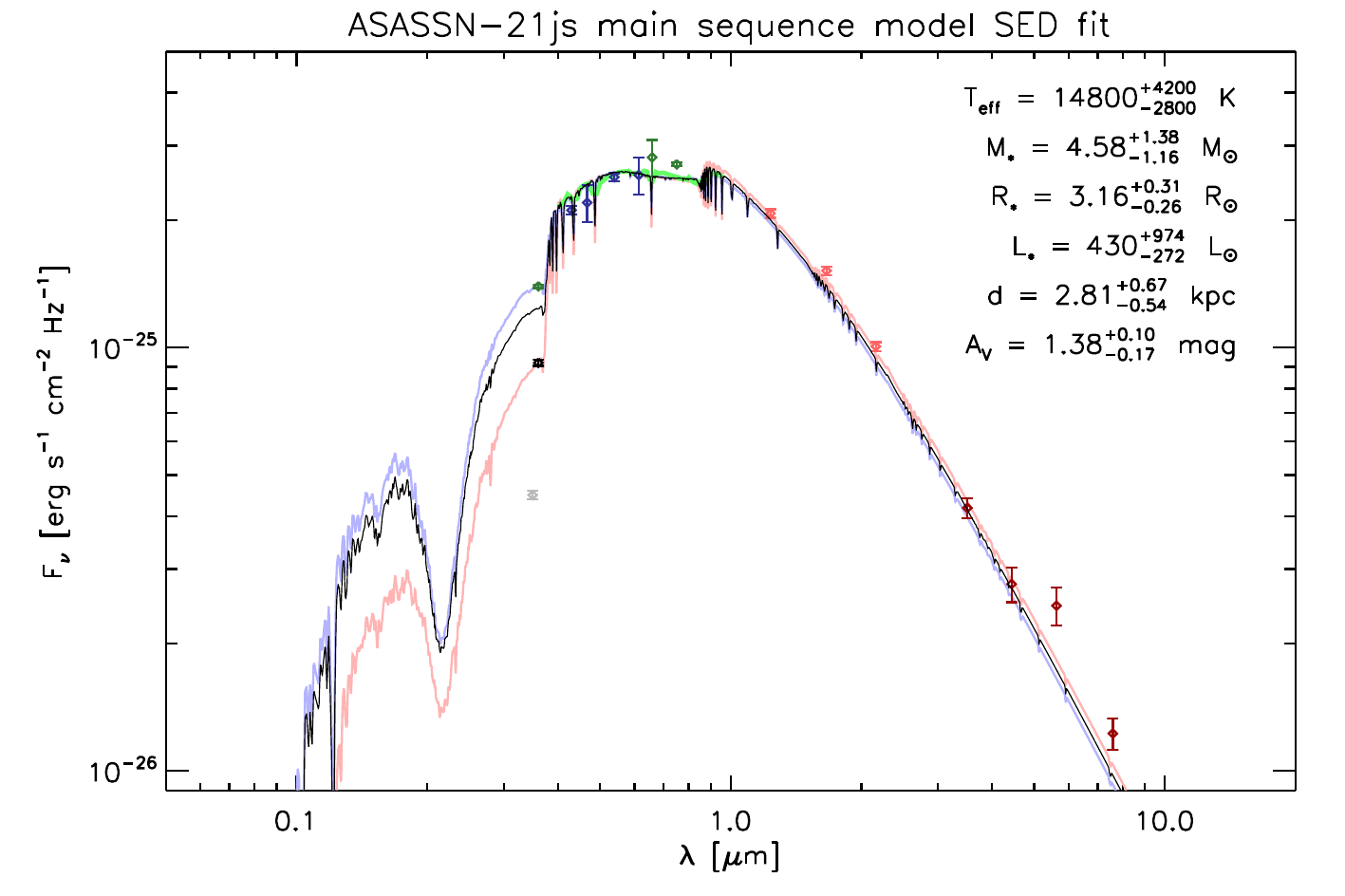}
    \caption{SED fit to the literature photometry and spectroscopy of ASASSN-21js. Data taken from (1) \cite{2014MNRAS.440.2036D}, (2) \cite{2002A&A...395..347E}, (3) \cite{2006AJ....131.1163S}, (4) \cite{2003PASP..115..953B,2009PASP..121..213C}, (5) \citet[][submitted to PASA]{2024arXiv240202015O}, (6) \cite{2023A&A...674A...1G}.}
    \label{fig:stellar_SED_fit}
\end{figure}

\newcommand{\Msun}{M$_\odot$}
\newcommand{\Teff}{$T_{\rm{eff}}$}
\newcommand{\Av}{$A_{\rm{V}}$}

In order to constrain the stellar properties of ASASSN-21js, we fitted a stellar model to the available literature photometry and the GAIA low-resolution spectrum.
Our model combines stellar evolutionary calculations with stellar model atmosphere calculations.
We assumed the star to be a main-sequence star with an age of one third of its nominal main-sequence lifetime.
We required the model to match the shape of the observed spectral energy distribution (SED), which yields the effective temperature and extinction estimate, as well as the absolute flux level, which yields the apparent angular diameter of the star.
For a given effective temperature and via our assumption about the main-sequence nature and the age of the object, the linear diameter of the star is dictated by the stellar evolutionary tracks, and hence the absolute flux level implies a photometric distance.
Because the quality of the GAIA parallax is good, we furthermore required that the inferred photometric distance match the value derived from the parallax. This effectively determines the effective temperature because the absolute brightness is directly related to the latter through stellar evolutionary models.

For stars with masses between 0.6 and 5.2~\Msun\, we used the Yonsei-Yale stellar evolutionary tracks \citep{2001ApJS..136..417Y,2002ApJS..143..499K,2003ApJS..144..259Y,2004ApJS..155..667D}, complemented by the tracks of \cite{1998A&A...337..403B} for stars below 0.6~\Msun \ and by those of \cite{2002A&A...391..195G} for stars above 5.2~\Msun.
At this point, we had a set of 50 models with effective temperatures from $\approx$2,000 monotonically increasing to $\approx$34,000~K, each with the appropriate stellar radius, mass, and age (the stellar luminosities and surface gravities are then implicitly given). We assumed solar metallicity throughout.
For each model, we calculated the emergent spectrum using PHOENIX models \citep{1997ApJ...483..390H,2013A&A...553A...6H} at effective temperatures below 10,000\,K and {\sc atlas\,9} models \citep{Kurucz1979,1992IAUS..149..225K,1994KurCD..19.....K} at higher temperatures.
To account for foreground extinction, we adopted the model by \cite{1989ApJ...345..245C} and assumed an interstellar reddening law (i.e. the total to the selective extinction ratio $R=3.1$).

We then proceeded to perform a fit to the SED, with the effective temperature, \Teff, and the visual extinction, \Av \ , as free parameters and with the requirement that the inferred photometric distance match the value derived from the GAIA parallax.
The resulting best fit is shown in Fig.~\ref{fig:stellar_SED_fit} with the black line. The model reproduces the SED well. In the temperature range in which we operate, the strength of the Balmer jump around 0.3645~\mum \ is a sensitive measure of effective temperature. We have two flux measurements blueward of the Balmer jump — that is, in the u-band — from \cite{2014MNRAS.440.2036D} and \cite{2024arXiv240202015O}, which differ by $\approx1.24$~mag. The model fit for the GAIA distance lies somewhat below the higher of the measured values, whereas there is no model that matches the lower of the measured values. We then proceeded to make a model fit that matches the higher u-band flux (light blue curve in Fig~\ref{fig:stellar_SED_fit}) and one that matches the average of the reported u-band fluxes (light red curve in Fig~\ref{fig:stellar_SED_fit}), in both cases letting go of the requirement to match the GAIA distance. We used these models to derive the envelope of the plausible range of parameters for the star, and derived our error estimates on the parameters reported in Table~\ref{table:asassn-21js-param} and Fig~\ref{fig:stellar_SED_fit} in this manner. In our further analysis, we used the parameters of the model that match that GAIA distance; that is, the black line in Fig.~\ref{fig:stellar_SED_fit}.

Photometry for ASASSN-21js is available from AAVSO, LCOGT, and TESS, and the star was detected at low signal-to-noise in WISE and the NEOWISE mission. 
Photometry from 2010 to 2021 does not show any significant variation, with measured magnitudes of $W1$ and $W2$.
Gaia photometry is available prior to the start of the eclipse, showing no significant photometric variation beyond what is seen in the ASAS-SN light curves presented here.
Due to the challenges of the heterogeneous nature of the these data, we do not carry out an analysis in this paper.

\section{The light curve}\label{sec:lightcurve}

We downloaded the photometry from the ASASSN web page, which contains both $V$ band and $g'$ band photometry.
The processing proceeded as follows: we first rejected bad points, which were flagged with a magnitude of 99.99.
Next, we rejected photometry with large error values: this was done by looking at the cumulative distribution function of the errors in the flux values of the $g'$ and $V$ filters and rejecting the top 10\% largest error bars.
Next was a visual inspection to remove single photometric outliers that had not been identified by the previous steps, informed by looking at the distribution of the photometry outside of the eclipse.

Both $g'$ and $V$ filter curves were normalized to the measured mean flux outside of the eclipse.
Therefore, normalization was done by first picking a time range when the datasets taken by both filters overlap outside of the eclipse, taking their respective mean values and normalizing the respective curves.
Finally, the data was binned with a bin width of nine days, which was a trade-off between increasing the signal-to-noise and minimising the computational cost of evaluating computer models.
Even though there are other possible astrophysical features contributing to variability, such as stellar spots and rotation \citep{balona_rotational_2016, balona_starspots_2017, sikora_mobster_2019}, their timescales are considerably shorter when compared to features in the observed eclipse.
The resulting light curve is plotted in Figure~\ref{fig:model_combine}.
A shallow eclipse (approximately 2\%) is now visible between MJD 58250 and MJD 59000, followed by the start of a clear deep eclipse starting just before MJD 59250 that has not finished to date.

\section{Ringed eclipser model}

After ruling out a planetary transit (the eclipse is too deep) or an eclipsing binary (the light curve is not symmetric), we hypothesise that the star is being transited by a foreground object hosting an azimuthally symmetric ring system, similar to J1407b \citep{2012AJ....143...72M, Kenworthy_2015}, PDS 110 \citep{Osborn17}, and Gaia21bcv \citep{Hodapp24}.
%,
The geometry and orientation of the ring system, as well as if the ring system is made up of multiple rings, will define the shape of the resultant light curve.
We therefore modelled the light curve with a computer-generated model of a limb-darkened star and an azimuthally symmetric ring system that was tilted and inclined to our line of sight \citep[using the {\tt exorings} package;][]{2015ascl.soft01012K}. 

\subsection{Determining basic parameters of the eclipser}
\label{sec:model_building}

We first determined a lower bound on the transverse velocity of a putative ring system by the measured gradients within the normalised flux curve.
Combined with an estimate of the diameter of the star, we determined the size of the eclipser along the chord made by the path of the star behind the ring, and determined an estimate of the orbital distance of the ring system from the star and the orbital period assuming a circular orbit.
We then estimated the initial geometry and parameters describing the ring system, so that these values could be used as a starting point for an algorithmic search to refine the model fit.
If we consider the transiting ring with a transmission, $T$, with a sharply defined ring edge that has a considerably larger diameter than the star, then the transit event can be simplified to be a straight edge occulter gradually eclipsing the star at a constant velocity, $v$, as is described in \cite{Kenworthy_2015}.

The speed at which the ring transits the star is proportional to the rate of change of the observed flux.
However, this will only constrain the obtained velocity, $v$, to be the minimum orbital velocity, $v_{\ensuremath{\mathrm{min}}}$, assuming that this gradient is the steepest, i.e. where the star's normalized flux,  $F'_{\mathrm{norm}}$, drops from 1 to 0 for the case where $T=1$.
The calculated $v_{\ensuremath{\mathrm{min}}}$ can be obtained through
\begin{equation}
\label{eq:min_v}
    v_{\mathrm{min}} = \frac{d}{t} = \frac{D_*}{|(1/m_{\mathrm{max}})|}
,\end{equation}
where $d$ is the distance traversed by the ring to completely eclipse the star, equalling the diameter of the star, $D_*$, $t$ is the time for $F'_{\ensuremath{\mathrm{ norm}}}$ to drop from 1 to 0, and $m_{\ensuremath{\mathrm{max}}}$ is the steepest observed gradient. 
We visually inspected the light curve to find the steepest gradient, which is approximated well with a straight line fit. %(see  Figure~\ref{fig:regression_fit}).
The steepest gradient obtained is between MJD 59201 and MJD 59278: $-(1.80\pm0.13)\times10^{-3}\,\text{day}^{-1}$.
Using the estimate of the stellar mass, Kepler's law, and assuming a circular orbit seen at high inclination, an orbital period ,$P$, orbital semi-major axis, $a_{\mathrm{orbit}}$, and diameter of the ring system, $D_{\mathrm{ring}}$, can be derived.
The previously obtained $v_{\ensuremath{\mathrm{min}}}$ suggests that $a_{\mathrm{orbit, \,max}} = (5.16\pm1.02)\times10^5\,\ensuremath{\mathrm{au}}$ or 2.5 pc. 

We then estimated the maximum period of the ring, $P_{\ensuremath{\mathrm{max}}}$, the predicted end of the transit, $t_{\ensuremath{\mathrm{end}}}$, the minimum diameter of the ring, $D_{\ensuremath{\mathrm{ring, \,min}}}$, and the minimum mass of the central object holding the ring, $m_{\ensuremath{\mathrm{obj, \,min}}}$.
We estimated the total time of the transit, $t_{\ensuremath{\mathrm{tot}}}$, by extrapolating the most recent photometry up to $F'_{\mathrm{ \,norm}}=1$.
This results in $t_{\ensuremath{\mathrm{end}}}=61046\pm 94$\ d and yields a total transit time of $t_{\ensuremath{\mathrm{tot}}}=2746\pm 94$\ d. 
All of these derived values are tabulated in Table~\ref{tab:derived_param_steep_grad}.
Limb darkening is the effect of the line of sight of the normal vector on a star's surface seen from the observer's point of view.
For this star, we assumed a linear limb-darkening model with $u=0.5$ consistent with this spectral type \citep{Sing10}, though we note that changing $u$ by 20\% does not significantly change the resultant models.

\subsection{Ring model}
\label{subsec:create_ring}

\begin{figure}[hbt]
\centering
    \includegraphics[width=\columnwidth]{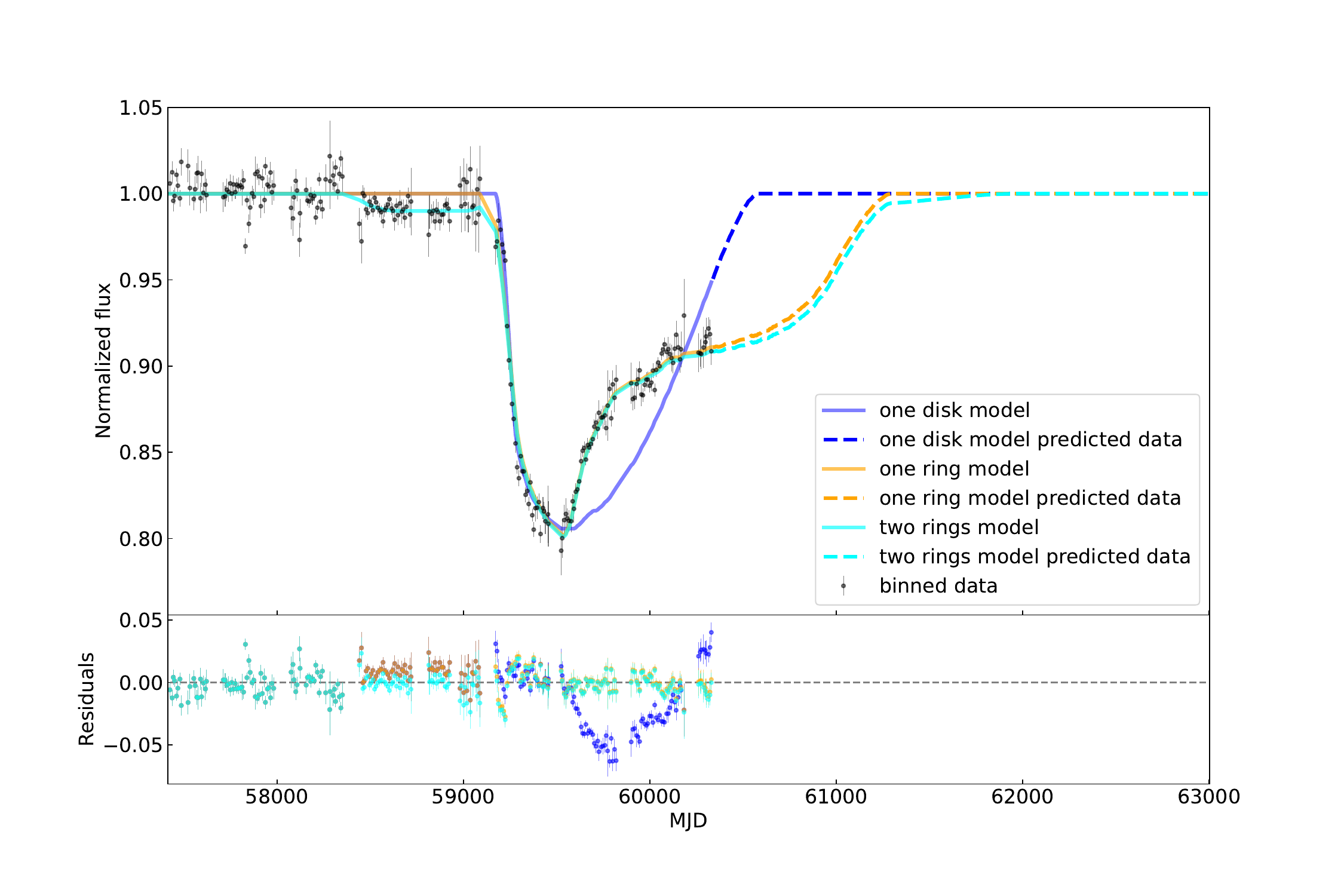}
    \caption{Plot of different models fitted to the data, which has been binned to 9 day intervals.}
    \label{fig:model_combine}
%    \script{plot_model_combine.py}
\end{figure}

To determine the geometry of the transiter, we started with a single inclined disc with a non-zero impact parameter, modelled according to the method in \citet{Kenworthy_2015}.
The modeling software assumes an origin at the centre of the disc, with the positive y axis due north on the sky and the positive x axis due east (PS=270 degrees).

Each disc has three parameters that define its geometry: the radius of each disc, $a_{\mathrm{disk}}$, its width, $w$, and transmission, $T$.
From an initially face-on and circular geometry, the model is then inclined to the plane of the sky by an inclination angle, $i$, and the resultant projected ellipse (with a semi-major axis parallel to the x axis) is then rotated anti-clockwise by an angle, $\phi$.
The star passes along a line parallel to the x axis and offset by a distance, $y_{\ensuremath{\mathrm{ring}}}$, with a velocity, $v$.
The $x$ position of the star at time $t$ is given by $x=v(t-t_0)$, where $t_0$ is the time when when the star crosses the y axis.
The resulting light curve model has been plotted against the real light curve and the $\chi^2$ evaluated to see how closely the model fits the data.

As can be seen in Figure~\ref{fig:model_combine}, a single tilted disc can fit the start of the deep eclipse between MJD 59200 and MJD 59500, but cannot fit the change in the light curve beyond the inflexion change at MJD 59500, with a $\chi^2=9500$.
If we take a disc with the same outer radius and geometry as the first model, but remove material at a smaller radius to change it to a ring, this model gives an almost perfect fit to the data after MJD 59000, but there is still a significant residual between MJD 58500 and MDJ 58590, with $\chi^2=1217$.
By adding a second ring model with the same inclination and tilt as the inner ring, we can then produce a fit at which the residuals are consistent with the noise in the stellar light curve and $\chi^2=738$.
We adopted the parameters of this two-ring model as a starting point for our MCMC fit.

\subsection{Applying MCMC}
\label{sec:apply_mcmc}

MCMC was used to perform sampling of the free parameters of the ring system to obtain posterior distributions of each parameter. 
This method also allowed us to obtain robust uncertainties of those sampled parameters and to observe the relations between each parameter; for example, how they correlate with each other and whether any degeneracy is present. 
Applying MCMC was done by using \texttt{emcee} \citep{emcee}. 
This Python package was specifically developed to apply MCMC to astrophysical problems and utilizes the affine-invariant ensemble sampler proposed by \cite{2010CAMCS...5...65G}.
% %
All parameters listed under the ``Free Parameters'' label from \ref{tab:best_fit} were explored throughout the run. 
All of the input values for the MCMC runs are listed in \ref{tab:emcee_all_fit}, while \ref{tab:emcee_prior} lists the prior distributions for each parameter. 
We ran for $2\times 10^6$ steps, discarded the first half of the chains, and thinned them by a factor of 100.

\begingroup
\renewcommand{\arraystretch}{1.5}
\begin{table}[tp]
\centering
\begin{threeparttable}
    \caption{Results from MCMC for the real dataset under the scenario with a system of two rings.}
\begin{tabular}{p{4cm} p{4cm}}    \hline
    \textbf{Parameter} & {\textbf{Value}}\\
    \hline
    \hline
    %Acceptance Rate & 0.080 \\
    %\hline
    \multicolumn{2}{c}{Free Parameters}\\    
    \hline
    $a_{\ensuremath{\mathrm{ring, \,outer}}} \, [\ensuremath{\mathrm{R_*}}]$ & $71.6_{-20.5}^{+28.9}$ \\
    $a_{\ensuremath{\mathrm{ring, \,inner}}} \, [\ensuremath{\mathrm{R_*}}]$ & $52.0_{-19.6}^{+27.9}$ \\
    $i \, [^\circ]$ & $87.18_{-0.62}^{+0.53}$ \\
    $\phi \, [^\circ]$ & $4.8_{-1.3}^{+1.4}$ \\
    $w_{\ensuremath{\mathrm{outer}}} \, [\ensuremath{\mathrm{R_*}}]$ & $19.0_{-4.3}^{+2.7}$\\
    $w_{\ensuremath{\mathrm{inner}}} \, [\ensuremath{\mathrm{R_*}}]$ & ${7.7}_{-0.9}^{+1.1}$ \\
    $T_{\ensuremath{\mathrm{outer}}}$ & $0.990_{-0.001}^{+0.001}$ \\
    $T_{\ensuremath{\mathrm{inner}}}$ & $0.73_{-0.07}^{+0.04}$ \\
    $y_{\ensuremath{\mathrm{ring}}} \, [\ensuremath{\mathrm{R_*}}]$ & $4.4_{-2.0}^{+3.5}$ \\
    $v \, [\ensuremath{\mathrm{D_*/ day}}]$ & $\left(1.1_{-0.1}^{+0.1}\right)\times10^{-2}$ \\
    $t_0 \, [\mathrm{day}]$ & $61464_{-845}^{+1203}$ \\
    \hline
    \multicolumn{2}{c}{Derived Values}\\
    \hline

    $m_{\ensuremath{\mathrm{obj}}} \, [\mathrm{M}_\oplus]$ & $\left ( 0.9_{-0.7}^{+2.3}\right ) \times 10^{-4}$ \\
    $P \, [\mathrm{yr}]$ & $\left ( 6.1_{-1.7}^{+2.5}\right )\times 10^5$  \\
    $a_{\mathrm{orbit}} \, [\ensuremath{\mathrm{R_*}}]$ & $\left(8.1_{-1.6}^{+2.1}\right)\times10^5$\\
    \hline
    \end{tabular}
    \label{tab:mcmc_result}
\end{threeparttable}
\end{table}
\endgroup

Looking at the comparison between the best manually fitted values (Table \ref{tab:best_fit}) and MCMC results (Table \ref{tab:mcmc_result}), they still exhibit differences, but these are not as drastic as previous comparisons.
This means that the best manually fitted model is a good approximation of the whole set of model solutions found by the algorithm in order to obtain the smallest $\chi^2$ value possible.
Overall, the solutions found through MCMC favour larger and faster ring systems.
A corner plot of the fitted distributions is shown in Figure~\ref{fig:two_ring_corner}, displaying thermalised and Gaussian distributions.

\begin{figure*}[tp]
    \centering
    \includegraphics[width=\linewidth]{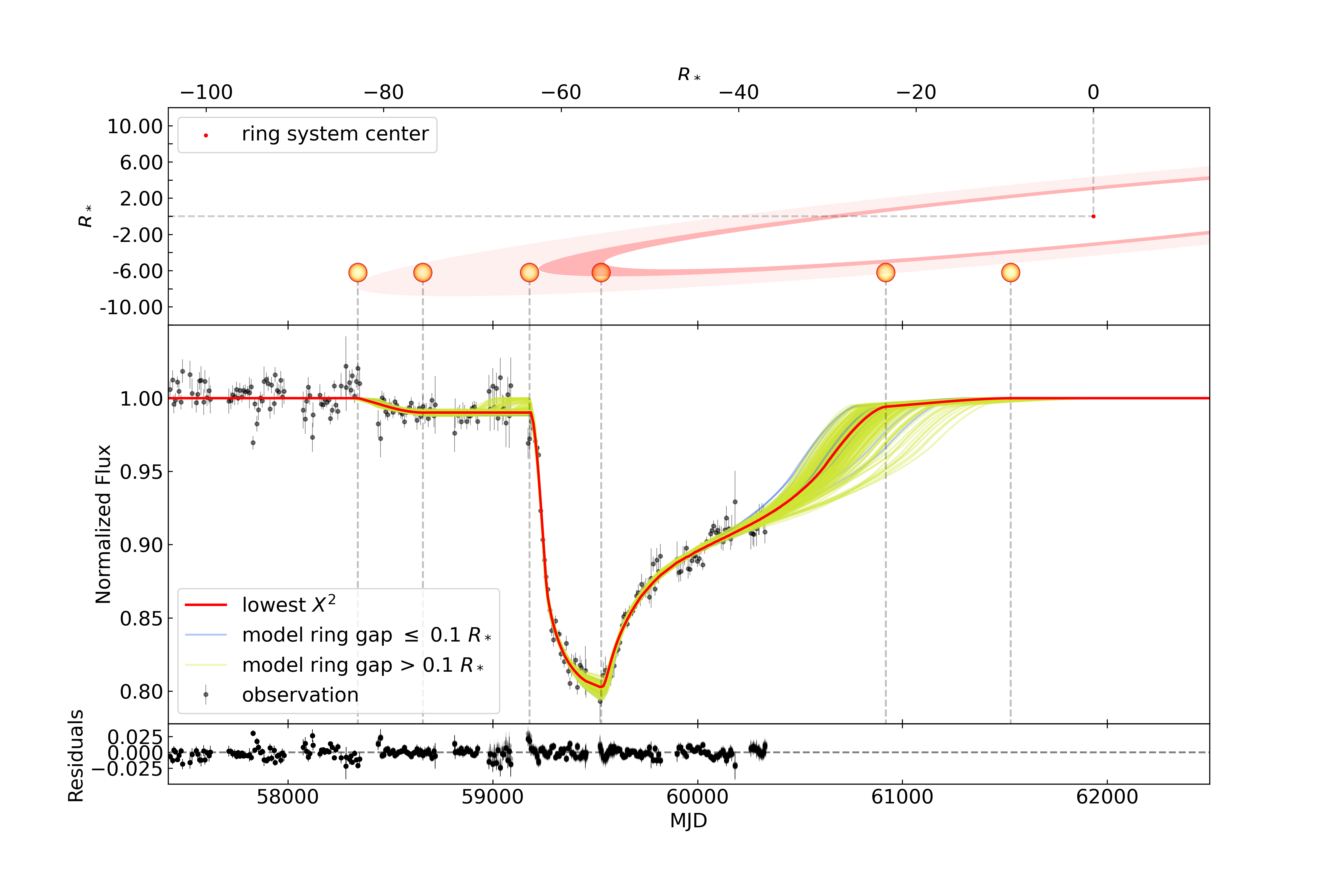}
    \caption{MCMC light curve models for the system with two rings and a gap.
   The figure on the topmost plot is based on the model with the lowest $\chi^2$ seen on the middle plot.
    We exaggerate the ring transmissions for better contrast.} 
    \label{fig:plot_lc_2ring_gap}
 %   \script{plot_vector_model_lc_resi_complete_mjd.py}
\end{figure*}

\section{Discussion}
\label{sec:Discussion}

When the star moves behind the leading edge of the ring system, the geometry causes most parts of the star to be eclipsed quickly, corresponding to the very steep ingress observed in the second transit. 
As the star progresses behind the system, the star emerges bit by bit behind the almost horizontal inner edge, resulting in the slow and gentle egress of the transit.
The additional outer ring can also recreate the first transit along with its asymmetric ingress and predicted egress. 
We also found that using the approach of optically thin rings with constant transmission is able to model the depth of the eclipse, minimising the number of free parameters required.
This has been observed in other systems hosting dusty rings like J1407b \citep{2012AJ....143...72M, Kenworthy_2015} and PDS 110b \citep{Osborn17}.
Running the MCMC shows two different distributions, one with a ring gap and one with no gap.
The greater proportion of the solutions favour no gap.
If we look at the other parameter values, they raise some questions about the physical interpretation of the system. 

Using the derived radius of the star, we can convert the model fit parameters into astrophysical units: the diameter of the inner ring is $1.53$ au, which in itself is not unusual -- J1407b and PDS~110b ring systems come to mind as similar diameter ringed structures around unseen companions.
The most notable feature of the model is that $v = 0.58$ \kms, corresponding to $a_{\mathrm{orbit}} \approx 12000$ au, placing this object within the Oort cloud of the central star.
Several planetary systems on wide orbits have been found. One of them is the COCONUTS-2 system \citep{2021ApJ...916L..11Z} comprising a $6.3\, \,{\mathrm{M_{jup}}}$ exoplanet orbiting an M3 class star at a projected distance of 6471 au. 
Since the obtained $a_{\mathrm{orbit}}$ value of ASASSN-21js is around twice that of COCONUTS-2's and also well under the maximum $a_{\mathrm{orbit}}$ possible derived from the steepest gradient, we consider the system to be bound.

We estimated the possible lower bound on the mass of the  companion by assuming that the rings fill out to one third of the Hill sphere of the companion. 
This gives a lower bound of $0.9\times 10^{-4} \mathrm{M_\oplus}$, which is on par with the masses of large asteroids and minor planets in the Solar System. 
A rough estimate of the masses of the rings can be made by assuming optically thin rings, particles of constant size and density.
For 1 micron radius particles with a density of 5 $\mathrm{g/cm}^3$, the mass of the rings is $5.6\times 10^{-6} \mathrm{M_\oplus}$ (4\% of the mass of Ceres), increasing to $5.6\times 10^{-3} \mathrm{M_\oplus}$ for 1 mm sized particles.
We note that this must underestimate the actual mass of the total ring system, since there will almost certainly be more material interior to the innermost projected distance probed by the path of the star behind the ring.
\section{Conclusions}

We hypothesise that the asymmetrical dips seen towards ASASSN-21js are caused by a transiting ring system rather than intrinsic variability of the star or a transit of non-ringed companion.
Further inspection of the light curve suggests that the transiting ring system might consist of two concentric rings with a total outer diameter of 2.11 au and that some of the models we identify have a clear gap between the two rings.
Additional data searches can also be performed for past observations of the system. 
We expect that, given the current transverse velocity, this will be the last eclipse we shall witness, but the discovery of other transits from archival data or future observations would contradict our derived long period for this system.
Rings around substellar objects in the Oort cloud of their parent stars may not be as unusual as they initially appear to be.
We note that four minor planets have had ring systems discovered around them and that Chariklo \citep{2014Natur.508...72B} notably has two rings around it, although these are orders of magnitude smaller than this system.

To assess the probability of observing this eclipse, assuming that the rings are around a companion bound to ASASSN-21js, we performed a Monte Carlo simulation using the radius and mass from Table 2.
We drew distributions of mass, radius, and transverse velocity for 500000 systems and calculated the probability of observing this ring eclipse over the lifetime of the ASAS-SN survey $\mathcal{O}$(10 years).
The probability is $1.0^{+0.5}_{-0.3}\times10^{-10}$ for one star.
We then repeated this calculation for a solar-type star with a mass and radius equal to the Sun, and the probability is $2.0^{+0.7}_{-0.4}\times10^{-12}$ for one star, so if we assume that ASAS-SN has observed $\mathcal{O}$(60 million stars) similar in mass and radius to the Sun, we obtain a probability of $1.2^{+0.4}_{-0.2}\times10^{-4}$ for all stars in the ASAS-SN survey (the probabilities quoted are for the 25\%, 50\%, and 75\% percentiles).
It seems therefore unlikely that we would observe such a transit in the ASAS-SN survey, unless we were seeing an unusually large and deep transit, or that many more and/or shallower transits are present in the ASAS-SN survey that have not yet been identified as such.

This transit is still ongoing and is predicted to last until MJD 61526 (May 1 2027), providing an opportunity for further observations to characterise the material making up the rings.
We would expect more multi-year transiting systems to be detected in the Gaia DR4 release, expected in 2026, and in the first few years of operation of the Vera C. Rubin telescope towards the end of 2025, enabling population statistics to be derived and limits on activity in the outer reaches of Oort clouds.

\begin{acknowledgements}

We thank the referee for their careful reading of our manuscript, which have helped improve our paper.
Part of this work was supported by the student fund from the Indonesia Endowment Fund for Education Scholarship (LPDP). 
This work has also made use of data taken from ASAS-SN \citep{shappee_man_2014, kochanek_all-sky_2017}, processed and obtained through \url{https://asas-sn.osu.edu/}.
This work has also made use of data from the European Space Agency (ESA) mission {\it Gaia} (\url{https://www.cosmos.esa.int/gaia}), processed by the {\it Gaia} Data Processing and Analysis Consortium (DPAC, \url{https://www.cosmos.esa.int/web/gaia/dpac/consortium}).
Funding for the DPAC has been provided by national institutions, in particular the institutions participating in the {\it Gaia} Multilateral Agreement. Data taken from tic v8.2 was accessed using the VizieR catalogue access tool, CDS, Strasbourg, France.%(DOI : 10.26093/cds/vizier). 
The original description of the VizieR service was published in 2000, A\&AS 143, 23 \citep{2000A&AS..143...23O}.
This publication makes use of VOSA, developed under the Spanish Virtual Observatory (\url{https://svo.cab.inta-csic.es}) project funded by MCIN/AEI/10.13039/501100011033/ through grant PID2020-112949GB-I00.
VOSA has been partially updated by using funding from the European Union's Horizon 2020 Research and Innovation Programme, under Grant Agreement nº 776403 (EXOPLANETS-A).

\end{acknowledgements}

\bibliographystyle{aa}
\bibliography{bib}

\begin{appendix}

\section{Order of magnitude estimate}

\begin{table}[ht]
    \centering
    \caption{Order of magnitude estimates for a ring system.}
    \begin{tabular}{lc}
        \hline
        \textbf{Parameter} & \textbf{Value}\\
        \hline
        \hline
        $v_{\ensuremath{\mathrm{min}}} \, [{\ensuremath{\mathrm{R_*/\ensuremath{\mathrm{day}}}}}]$ & $(3.63\pm0.38)\times10^{-3}$\\
        $a_{\ensuremath{\mathrm{orbit, \,max}}} \,[{\ensuremath{\mathrm{R_*}}}]$ & $(3.00\pm0.69)\times10^7$\\
        $P_{\ensuremath{\mathrm{max}}} \, [\ensuremath{\mathrm{yr}}]$ & $(1.42\pm0.40)\times10^8$\\
        $t_{\ensuremath{\mathrm{end}}} \, [\ensuremath{\mathrm{day}}]$ & $61046.65\pm94.20$\\
        $D_{\ensuremath{\mathrm{disk, \,min}}} \,[{\ensuremath{\mathrm{R_*}}}]$ & $9.97\pm1.11$\\
    %    $m_{\ensuremath{\mathrm{obj, min}}} \, [M_\oplus]$ & $(1.76\pm1.85)\times10^{-12}$\\
        \hline
    \end{tabular}
    \label{tab:derived_param_steep_grad}
\end{table}

\section{Best manual fitting parameters}

\begin{table}[ht]
\centering
\begin{threeparttable}
    \caption{Best manual result for two ring system.}
    \begin{tabular}{p{3cm} p{3cm}}
    \hline
    \textbf{Parameter} & {\textbf{Value}}\\
    \hline
    \hline
    \multicolumn{2}{c}{Fixed Parameters}\\
    \hline
    $R_{\ensuremath{\mathrm{star}}} \, [\ensuremath{\mathrm{px}}]$ & 50 \\
    $u$ & 0.5 \\
    \hline
    \multicolumn{2}{c}{Free Parameters}\\
    \hline
    $a_{\ensuremath{\mathrm{ring, \,outer}}} \, [\ensuremath{\mathrm{R_*}}]$ & $15.6{\pm2.2}$ \\
    $a_{\ensuremath{\mathrm{ring, \,inner}}} \, [\ensuremath{\mathrm{R_*}}]$ & $9.0{\pm1.3}$ \\
    $i \, [^\circ]$ & $81.5$ \\
    $\phi \, [^\circ]$ & $7.5$ \\
    $w_{\ensuremath{\mathrm{outer}}} \, [\ensuremath{\mathrm{R_*}}]$ & $4.10{\pm0.6}$ \\
    $w_{\ensuremath{\mathrm{inner}}} \, [\ensuremath{\mathrm{R_*}}]$ & $2.70\pm0.4$ \\
    $T_{\ensuremath{\mathrm{outer}}}$ & $0.99$ \\
    $T_{\ensuremath{\mathrm{inner}}}$ & $0.66$ \\
    $y_{\ensuremath{\mathrm{ring}}} \, [\ensuremath{\mathrm{R_*}}]$ & $1.60 \pm0.22$ \\
    $v$ [\ensuremath{\mathrm{D_*}}/day] & $0.0035\pm0.0002$ \\
    $t_0 \, [\text{day}]$ & $60480.801$ \\
    \hline
    \multicolumn{2}{c}{Derived Values}\\
    \hline
    $X^2$ & $764.55$ \\
    $t_{\ensuremath{\mathrm{end}}} \, [\text{day}]^a$ & $61869.206$ \\
    $m_{\ensuremath{\mathrm{obj}}} \, [\mathrm{M}_\oplus]$ & $(0.7\pm0.8)\times 10^{-9}$ \\
    $P \, [\text{yr}]$ & $(2.1\pm0.9)\times 10^7$  \\
    $a_{\mathrm{orbit}} \, [\ensuremath{\mathrm{R_*}}]$ & $(1.3\pm0.5)\times10^5$ \\
    \hline
    \end{tabular}
     \tablefoot{
     \tablefoottext{a}{Based on extrapolating the model past the last available observed data point by increments of $t = 9$ days until $F'_{\ensuremath{\mathrm{norm}}} = 1$.}
}        
    \label{tab:best_fit}
\end{threeparttable}
\end{table}

\FloatBarrier
\newpage
\section{Parameters for emcee}

%\begin{landscape}
\begin{table}[ht]
\centering
\begin{threeparttable}
    \caption{\texttt{emcee} inputs for two ring system.}
    \begin{tabular}{p{3cm} p{3cm}}
    \hline
    \textbf{Parameter} & {\textbf{Value}}\\
    \hline
    \hline
    \multicolumn{2}{c}{\texttt{emcee} Parameters}\\
    \hline
    $n_{\ensuremath{\mathrm{dim}}}$ & 11 \\
    $n_{\ensuremath{\mathrm{walkers}}}^a$ & 40  \\
    $n_{\ensuremath{\mathrm{chain}}}$ & $2\times10^6$ \\
    Move$^b$ & Stretch Move\\
    \hline
    \multicolumn{2}{c}{System Parameters}\\
    \hline
    $a_{\ensuremath{\mathrm{ring, \,outer}}} \, [\ensuremath{\mathrm{px}}]$ & 780 \\
    $a_{\ensuremath{\mathrm{ring, \,inner}}} \, [\ensuremath{\mathrm{px}}]$ & 450.8 \\
    $i \, [^\circ]$ & 81.5 \\
    $\phi \, [^\circ]$ &  7.5\\
    $w_{\ensuremath{\mathrm{outer}}} \, [\ensuremath{\mathrm{px}}]$ & 205 \\
    $w_{\ensuremath{\mathrm{inner}}} \, [\ensuremath{\mathrm{px}}]$ & 135.5 \\
    $T_{\ensuremath{\mathrm{outer}}}$ & 0.99 \\
    $T_{\ensuremath{\mathrm{inner}}}$ & 0.66 \\
    $y_{\ensuremath{\mathrm{ring}}} \, [\ensuremath{\mathrm{px}}]$ & 80 \\
    $v$ [\ensuremath{\mathrm{px}}/day] & 0.35 \\
    $t_0$ [day] & 60480.801 \\
    \hline
    \end{tabular}
 \tablefoot{
      \tablefoottext{a}{\cite{emcee} suggests that $n_{\ensuremath{\mathrm{walkers}}} \geq 3\,n_{\ensuremath{\mathrm{dim}}}$ or, even better, $n_{\ensuremath{\mathrm{walkers}}} \gg n_{\ensuremath{\mathrm{dim}}}$. The use of $n_{\ensuremath{\mathrm{walkers}}} = 40 > n_{\ensuremath{\mathrm{dim}}}$ is sufficient to compensate speed for performance issue. Using more $n_{\ensuremath{\mathrm{walkers}}}$ was seen to decrease the running time significantly.}
    \tablefoottext{b}{The default move in \texttt{emcee} based on \cite{2010CAMCS...5...65G}.}
    }
    \label{tab:emcee_all_fit}
\end{threeparttable}
\end{table}
\newpage
\FloatBarrier

\section{Prior distributions for emcee}

\begin{table}[ht]
\centering
\begin{threeparttable}
    \caption{Prior distributions for \texttt{emcee} parameters.}
    \begin{tabular}{lll}
    \hline
    \textbf{Parameter} & {\textbf{Boundary}} & \textbf{Dist.}\\
    \hline
    \hline
    $a_{\ensuremath{\mathrm{ring, \,outer}}}$ & $a_{\ensuremath{\mathrm{ring, \,outer}}} > 0$ & Uniform \\
    $a_{\ensuremath{\mathrm{ring, \,inner}}}$ & $0 < a_{\ensuremath{\mathrm{ring,\,inner}}} < (a_{\ensuremath{\mathrm{ring, \,outer}}}-w_{\ensuremath{\mathrm{outer}}})$ & Uniform \\
    $i$ & $0^\circ \leq i < 90^\circ$ & Uniform \\
    $\phi^a$ & $0^\circ \leq \phi \leq 180^\circ$ & Uniform \\
    $w_{\ensuremath{\mathrm{outer}}}$ & $0 < w_{\ensuremath{\mathrm{outer}}} < a_{\ensuremath{\mathrm{ring, \,outer}}}$ & Uniform \\
    $w_{\ensuremath{\mathrm{inner}}}$ & $0 < w_{\ensuremath{\mathrm{inner}}} < a_{\ensuremath{\mathrm{ring, \,inner}}}$ & Uniform \\
    $T_{\ensuremath{\mathrm{outer}}}$ & $0 < T_{\ensuremath{\mathrm{inner}}} < 1$ & Uniform \\
    $T_{\ensuremath{\mathrm{inner}}}$ & $0 < T_{\ensuremath{\mathrm{inner}}} < 1$ & Uniform \\
    $y_{\ensuremath{\mathrm{ring}}}$ & $ y_{\ensuremath{\mathrm{ring}}} \geq 0$ & Uniform \\
    $v$ & $v > 0$ & Uniform \\
    $t_0^b$ & $58000 \leq t_0 \leq 64000$ & Uniform \\
    \hline
    \end{tabular}
 \tablefoot{
      \tablefoottext{a}{Although technically $\phi$ has a range of $0^\circ \leq \phi \leq 360^\circ$, the ring system, which due to having a value for $i$ is drawn as an ellipse, has a rotational symmetry when it is rotated $180^\circ$, making evaluating for only the 1st and 2nd quadrant sufficient.}
      \tablefoottext{b}{
Intuitively, $t_0$ cannot be passed before the start of the dip.
      Because there are two dips and the second dip is longer than the first, it is safe to say that it will also not be passed before the start of the second dip, hence the lower boundary.
      The upper boundary is also intuitively should not be lower than the predicted $t_{\ensuremath{\mathrm{end}}}$. 
      To be on the safe side, it is set to a much higher value than $t_{\ensuremath{\mathrm{end}}}$ so that a good amount of values when $t_0 > t_{\ensuremath{\mathrm{end}}}$ can also be accommodated in the MCMC.}}
    \label{tab:emcee_prior}
\end{threeparttable}
\end{table}

\clearpage
\FloatBarrier

\section{Corner plot for emcee fit}

\begin{figure}[h]
    \centering
    \includegraphics[width=2.0\linewidth]{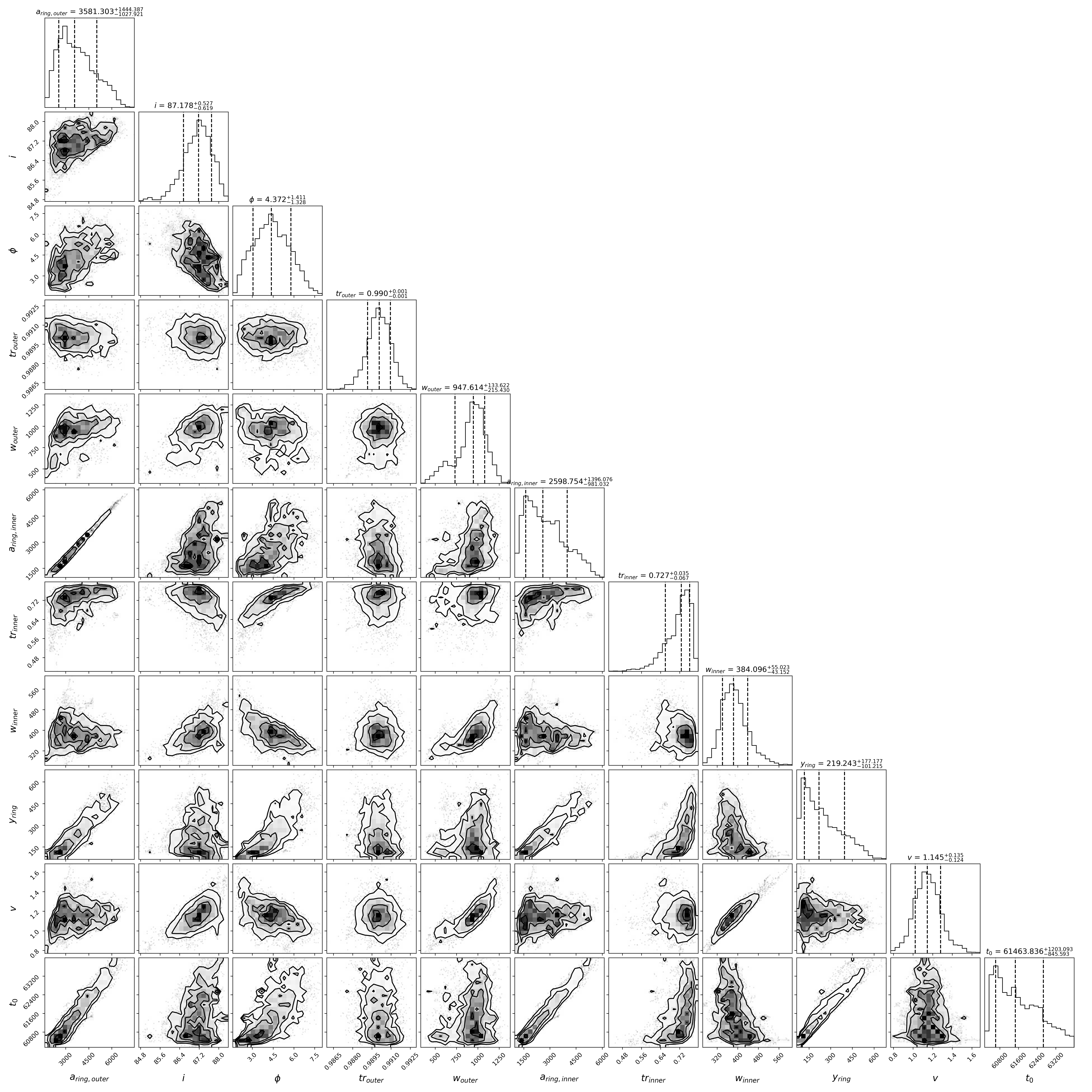}
    \caption{Corner plot for the two ring model system.} 
    \label{fig:two_ring_corner}
 %   \script{plot_vector_model_lc_resi_complete_mjd.py}
\end{figure}

\end{appendix}

\end{document}